\documentclass{article}
\usepackage{spconf,amsmath,graphicx}
\usepackage{booktabs}
\usepackage{mathalpha}
\usepackage{siunitx}
\usepackage{amssymb}
\usepackage{hyperref}
\usepackage{enumitem}
\usepackage{cite}
\usepackage{multirow}
\usepackage{etoolbox,siunitx}
\robustify\bfseries
\usepackage{booktabs}
\usepackage{balance}

\title{Why does music source separation benefit from cacophony?}
\name{Chang-Bin Jeon$^{1,2}$, %
      Gordon Wichern$^{1}$,
      François G. Germain$^{1}$,
      Jonathan Le Roux$^{1}$\thanks{This work was performed while C.-B.~Jeon was an intern at MERL.}}
\address{$^1$Mitsubishi Electric Research Laboratories (MERL), Cambridge, MA, USA\\
$^2$Department of Intelligence and Information, Seoul National University, Seoul, South Korea}
\begin{document}
\ninept
\maketitle
\setlength{\abovedisplayskip}{3pt}
\setlength{\belowdisplayskip}{3pt}
\begin{abstract}
In music source separation, a standard training data augmentation procedure is to create new training samples by randomly combining instrument stems from different songs. These random mixes have mismatched characteristics compared to real music, e.g., the different stems do not have consistent beat or tonality, resulting in a cacophony. In this work, we investigate why random mixing is effective when training a state-of-the-art music source separation model in spite of the apparent distribution shift it creates. Additionally, we examine why performance levels off despite potentially limitless combinations, and examine the sensitivity of music source separation performance to differences in beat and tonality of the instrumental sources in a mixture.

\end{abstract}
\vspace{-.3cm}
\section{Introduction}
\label{sec:intro}
\vspace{-.1cm}

Music source separation (MSS) has seen tremendous progress in performance in the deep learning era with an ever-growing list of high-performing models, in both the open-source community and commercial applications~\cite{stoter2019open, hennequin2020spleeter}. While this success is often credited to advances in network architecture~\cite{takahashi2018mmdenselstm, defossez2021hybrid, luo2023music} and growing public dataset availability~\cite{MUSDB18HQ, manilow2019cutting, pereira2023moisesdb}, the contribution of training pipeline improvements (e.g., data augmentation) has gone largely unnoticed and remains poorly understood. Meanwhile, MSS research continues to suffer from a chronic lack of available large-scale training data. Indeed, despite some efforts to leverage self-supervised training \cite{chen2023pachubert}, training a state-of-the-art MSS model still requires access to large amounts of supervised data for which the stems (i.e., recordings of the isolated sources) that make up a song are available for use as training targets. This remains difficult due to the practical barriers to recording them (especially outside of a recording studio environment) and intellectual property concerns. The size of MSS datasets thus tends to be small, with MUSDB18~\cite{MUSDB18HQ}, the most widely used dataset in the field, containing only 150 individual songs, i.e., about \num{10} hours per stem.

Data augmentation techniques have long been used to improve performance, especially in the context of low training data availability. Among these, %
techniques relying on random combinations of samples have become popular in many fields, thanks to their simplicity and effectiveness. For example, approaches such as mixup~\cite{zhang2018mixup} or cutmix~\cite{yun2019cutmix} have been successfully applied to the training of various classifier systems, including audio ones~\cite{kong2020panns}. Random mixing, in the sense of mixing together randomly selected audio sources, is also a staple of the learning-based audio source separation literature, from speech enhancement and separation~\cite{WDL2018, Hershey2016} to natural sound separation~\cite{kavalerov2019universal} and soundtrack separation~\cite{petermann2022cfp}. It is however important to note that, in the context of these applications, a random mix generally results in somewhat plausible sound scenes. %
One could thus argue these mixes remain conceptually ``in-distribution'' with respect to the target task so that the benefit obtained from their use at training should feel unsurprising.

Crucially, we also find random mixing in much of the MSS literature, where new ``songs'' are obtained by randomly combining stems from different songs into additional training mixtures~\cite{uhlich2017improving}. Thanks to its consistent effectiveness, the technique has become standard practice when training MSS networks, especially in the typical 4-stem (vocals, bass, drums, and other) separation setting found in recent challenges~\cite{stoter2018, mitsufuji2021mdx, fabbro2023sound}. However, unlike for many of the other fields mentioned earlier, the basis for the success of random mixing is less clear in the MSS context. Indeed, the target audio for the task is characterized by the strong consistency between the stems of a given song, as they are made to fit together in terms of aspects like beat, key, timbre, etc. Randomly-mixed mixtures exhibit no such consistency and are akin to a cacophony, making them conceptually ``out-of-domain'' with respect to the target task and, as such, potentially detrimental to the training of better-performing systems. 
Similarly, we lack a solid justification regarding why random mixing was helpful in a recent paper on multiple singing voices separation \cite{jeon2023medleyvox}, where sources are arguably even more consistent with other sources (e.g., in timbre, pitch, and time).

Data augmentation approaches were proposed to explicitly reduce this domain mismatch, either as replacement of or as complement to random mixing. For example, generating synthesized stems out of the many publicly-available MIDI songs~\cite{manilow2019cutting} allows generation of many new diverse consistent mixtures, while deploying time stretching and pitch shifting in combination with random mixing~\cite{defossez2021hybrid} allows creation of more consistent mixtures, e.g., in key and beat. However, in spite of their sensible foundations, these techniques have yet to demonstrate consistent gains over using random mixing alone. Here, we aim at elucidating some of the reasons behind the success of random mixing, focusing on the natural question: \textit{what do separation networks learn from collections of unmusical random mixes?}
In particular, we raise the following research questions:
\begin{itemize}[leftmargin=*]\setlength\itemsep{0mm}
  \item {Why do networks trained with randomly mixed training examples perform better than ones trained without random mixing in spite of the distribution shift?}
  \item {Random mixing can make a limitless amount of training data but performance gains level off. To which extent can random mixing give us a benefit? }
  \item {In our intuition, key differences between random and original mixes are whether they have consistent beat (or time) and tonality across each stem. How do these characteristics affect MSS?}
\end{itemize}

To answer these, we compare the training dynamics of networks with and without using random mixing. Next, we investigate the effective number of unique random mixes in training. We also study how inconsistent beat or tonality among the different instruments in a mix affect performance through evaluations on pitch- or time-shifted test data. Lastly, we confirm that pitch shift and slight time shift can lead to a performance gain without random mixing, even if less than when combined with it.

In this paper, we focus on the 4-stem separation setting based on MUSDB-HQ data, where stems may be strongly correlated in pitch or time but generally have very different timbral characteristics. Although we leave analysis on different MSS tasks as future work, we believe our results and conclusions can shed light on understanding random mixing in diverse separation tasks.

\vspace{-.2cm}
\section{music source separation}
\vspace{-.1cm}
\subsection{Task Definition}
The task of music source separation consists in extracting the signal of individual sources (typically referred to as \textit{stems}) from a musical input mixture. Mathematically, we consider a mixture signal $x$ as the sum of $N$ stems $x^{(n)}$, i.e.,
\begin{equation} \label{eqn:mss}
    x = \sum_{n=1}^{N} x^{(n)}.
\end{equation}
The task consists then in putting together a system (e.g., an MSS model) which, from input $x$, generates its best estimate $\hat{x}^{(n)}$ of $x^{(n)}$ for each $n$. We note that the definition of ``source'' is ambiguous, and musical songs include different numbers and types of instruments. Subsequently, we adopt the simplifying convention from the typical 4-stem MSS task \cite{stoter2018, mitsufuji2021mdx, fabbro2023sound} where we define sources as belonging to a fixed closed set of $N\!=\!4$ distinct source types, i.e., \textit{vocals, drums, bass}, and \textit{other}. The signal from a given musical instrument is then found in the stem of its corresponding source type. We also adopt the typical convention that all signals are stereo (i.e., 2-channel) signals.

\vspace{-.2cm}
\subsection{Experimental Setup}
\noindent{\bf TFC-TDF-UNet v3:}
To make our analysis as current as possible, we use the state-of-the art TFC-TDF-UNet v3 architecture \cite{kim2023sound}, winner of the recent 2023 MDX Challenge Leaderboard A \cite{fabbro2023sound}. %
The model takes as input the short-time Fourier transform (STFT) of the mixture signal $x$, outputs an estimated STFT for all sources, and is trained using mean-squared error loss between the waveforms of the estimated stems (obtained as the inverse STFT of the network output) and their ground truths. 
In our experiments, we use the same TFC-TDF-UNet v3 settings and hyperparameters that were used to train the model on the MUSDB-HQ dataset in the original paper, including the use of 6-second training chunks, overlap-add at inference time, model exponential moving average (EMA) \cite{polyak1992acceleration}, and automatic mixed precision training \cite{micikevicius2018mixed}. We only change the batch size to 4 to fit the model on a single 24 GB memory GPU. Every model instance was trained for 470k steps. %
Additionally, we employ loudness normalization \cite{jeon2022towards} to minimize the domain mismatch in terms of loudness. %

\noindent{\bf MUSDB-HQ}~\cite{MUSDB18HQ} consists of 150 songs split into a 100 song training set and a 50 song test set (3.5 hours). 
Following common practice, we divide the training set using 86 songs for training (5.3 hours) and 14 for validation (1.0 hours).

\noindent{\bf Evaluation Metric:} We use the standard source-to-distortion ratio (SDR) \cite{vincent2006performance} as implemented in museval \cite{stoter2018}. However, for the few experiments (Fig.~\ref{fig:eval_on_train}) that require hundreds of evaluations for every 10k training steps, we use instead the more computationally efficient SDR metric proposed at the recent SDX challenge \cite{fabbro2023sound}, denoted as $\overline{\text{SDR}}$. Except in Fig.~\ref{fig:randmix_ratio}, all scores are averaged across the 4 stems.

\noindent{\bf Data Augmentation:} Since music with stem data is highly difficult to acquire, researchers have used various data augmentation methods to maximize the potential of small public datasets. %
As described in the introduction,  the focus of this paper is random mixing augmentation~\cite{uhlich2017improving}. During the sampling of a training data sample, we dynamically generate a random mixture $y$ as follows, where we denote the $m$-th song in the dataset as $s_m$: for each source type $n$, we randomly select a song index $m_n$ and a start time $t_n$ and define the stem $y^{(n)}$ as the 6-second excerpt starting at time $t_n$ of the $n$-th stem $s_{m_n}^{(n)}$ of song $s_{m_n}$, and we add these $N$ stems together, i.e.,
\begin{equation} \label{eqn:rand_mix}
    y(t) = \sum_{n=1}^{N} y^{(n)}(t) = \sum_{n=1}^{N} s_{m_n}^{(n)}(t+t_n).
\end{equation}
This enables the creation of unique input random mixtures at each training iteration, even with a small amount of stem data. We denote as \emph{original mixes} the samples obtained from the original fixed mixtures without using random mixing from~\eqref{eqn:mss}, and as \emph{random mixes} the randomly sampled mixtures from~\eqref{eqn:rand_mix}. 
Additionally, we use pitch-shift data augmentation \cite{cohen2019improving} on all training mixes to best match the state-of-the-art configuration in \cite{kim2023sound}.

\section{Should we include original mixes at all?}
We experiment with various combinations of original and random mixes during model training to investigate how original mixes contribute to a network's final performance. Specifically, we set a probability $p$ for each experiment such that each mix in a batch of training data is sampled as an original mix with probability $p$ and as a random mix with probability $1\!-\!p$.

\begin{figure}[t]

\begin{minipage}[b]{1.0\linewidth}
  \centering
  \centerline{\includegraphics[width=7.6cm]{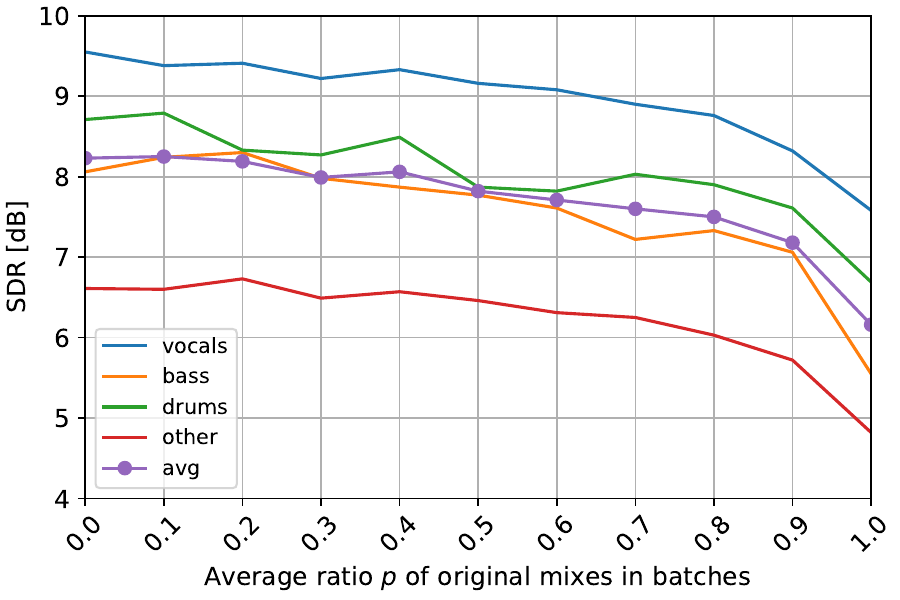}}
\end{minipage}
\vspace{-0.7cm}
\caption{%
Test SDR for networks trained with different ratios of original and random mixes. 
`$p\!=\!1.0$' corresponds to the model trained solely with original mixes and `$p\!=\!0.0$' to the model trained solely with random mixes.
}
\vspace{-.4cm}
\label{fig:randmix_ratio}
\end{figure}

When random mixing was first used in the context of music source separation \cite{uhlich2017improving}, the performance gain was found to be only \SI{0.2}{\dB} SDR (BLSTM-1 in Table 2 of \cite{uhlich2017improving}). However, we observe a \SI{2.1}{\dB} gain when training with only random mixes ($p\!=\!0.0$) compared to training with only original mixes ($p\!=\!1.0$) (see Fig.~\ref{fig:randmix_ratio}). We contend that this is due to a much more expressive model architecture: TFC-TDF-UNet v3 is a deep U-Net with 70M parameters, whereas BLSTM-1 is a shallow recurrent network with a single BLSTM layer. %
This allows our model to learn more diverse features from the infinitely many augmented mixes.
We also observe that the $p\!=\!0.1$ model does not show significant difference with the $p\!=\!0.0$ model (\SI{0.02}{\dB}). %
This implies that when we train models, we do not really need to use original mixes at all.

Interestingly, in Fig.~\ref{fig:randmix_ratio}, using random mixes even with only a probability of 0.1 ($p\!=\!0.9$) results in a substantial performance gain of \SI{1.0}{\dB}. Since we train models with batch size $B\!=\!4$ for $I\!=\!470$k iterations, a total of \num{1880000} different  mixes are generated over the duration of the training. That is, the $p\!=\!0.9$ model has learned significantly meaningful information from only \num{188000} randomly mixed training examples. This leads to the necessity for further investigations on the effective number of random mixes, which we will discuss in Section \ref{sec:effective}.

\subsection{Training Dynamics Comparison}

\begin{figure}[t]
\begin{minipage}[b]{0.48\linewidth}
  \centering
  \centerline{\includegraphics[width=4.0cm]{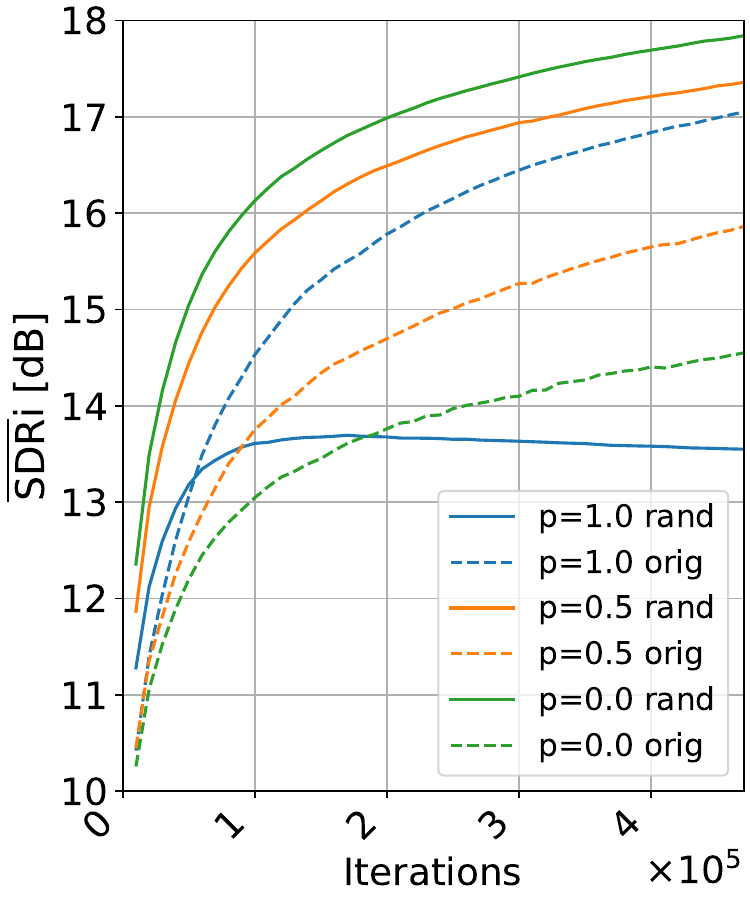}}
  \centerline{(a) Training set mixes}\medskip
\end{minipage}
\hfill
\begin{minipage}[b]{0.48\linewidth}
  \centering
  \centerline{\includegraphics[width=4.0cm]{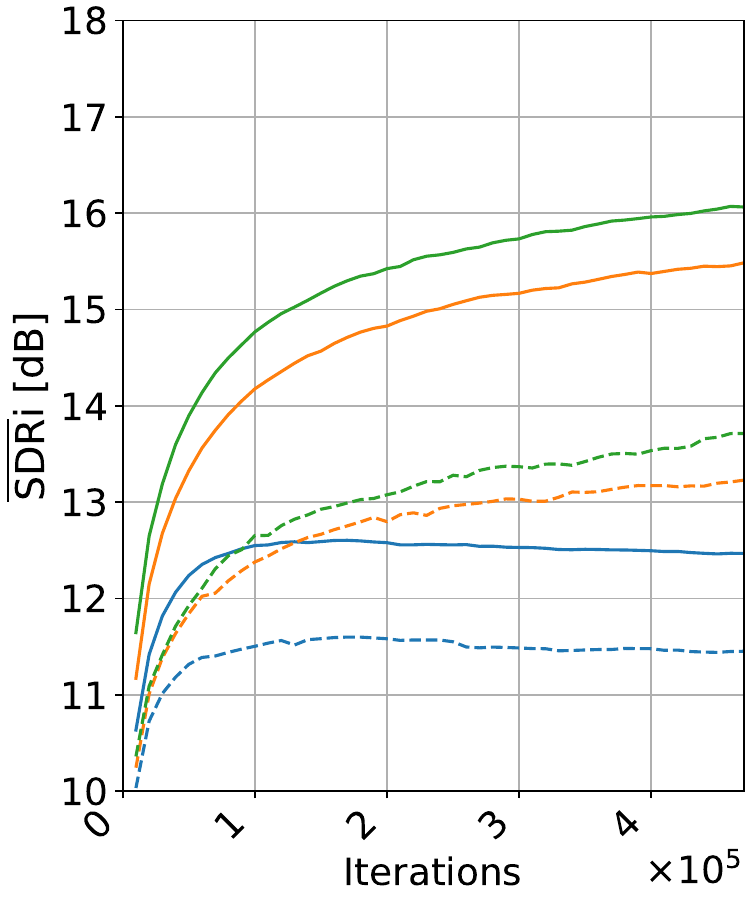}}
  \centerline{(b) Validation set mixes}\medskip
\end{minipage}
\vspace{-0.5cm}
\caption{%
Average $\overline{\text{SDR}}$ improvement during training as measured on sets of random (solid lines) and original mixes (dashed lines) for three different models. The model trained with only original mixes ($p\!=\!1.0$) overfits to the training data. Other models trained with random mixes ($p\!=\!0.5$ and $p\!=\!0.0$) do not.}
\label{fig:eval_on_train}
\end{figure}

In Fig.~\ref{fig:eval_on_train}, we compare the performance of three models ($p\in\{1.0, 0.5, 0.0\}$) over the course of training, measured every 10k iterations. 
Performance is evaluated on four sets consisting of either original mixes or random mixes, generated from either training or validation data.
These sets are generated once and fixed across training iterations. To match the training setting, evaluations are performed on 6-second chunks (for original mixes, songs are split without overlap).
For each stem's evaluation, random mixes are generated independently across 3 different seeds to marginalize the influence of randomness, resulting in twelve times as many chunks as the original mixes.

Noticeably, for the model only trained with original mixes ($p\!=\!1.0$) shown in the blue curves in Fig.~\ref{fig:eval_on_train}, the scores on original mixes in the training set are increasing while the scores on random mixes in the training set and both conditions in the validation set are decreasing. That is, the model trained without using random mixes easily overfits to the training data. On the other hand, models trained with at least some random mixes ($p\!=\!0.5$ and $p\!=\!0.0$, i.e., orange and green curves, respectively) do not appear to overfit.

In Fig.~\ref{fig:eval_on_train}(a), we also observe that the $p\!=\!0.5$ model (dashed orange curve) performs better than the $p\!=\!0.0$ model (dashed green curve) on the training original mixes, while performing worse on the validation original mixes in Fig.~\ref{fig:eval_on_train}(b). %
We conjecture this is because the $p\!=\!0.5$ model has seen the same mixes too many times during training, which we explore in more detail in Section \ref{sec:effective}.  %

\vspace{-.2cm}
\subsection{Influence of the Number and Variety of Random Mixes} \label{sec:effective}

\begin{figure}[t]
\begin{minipage}[b]{.48\linewidth}
  \centering
  \centerline{\includegraphics[width=4.0cm]{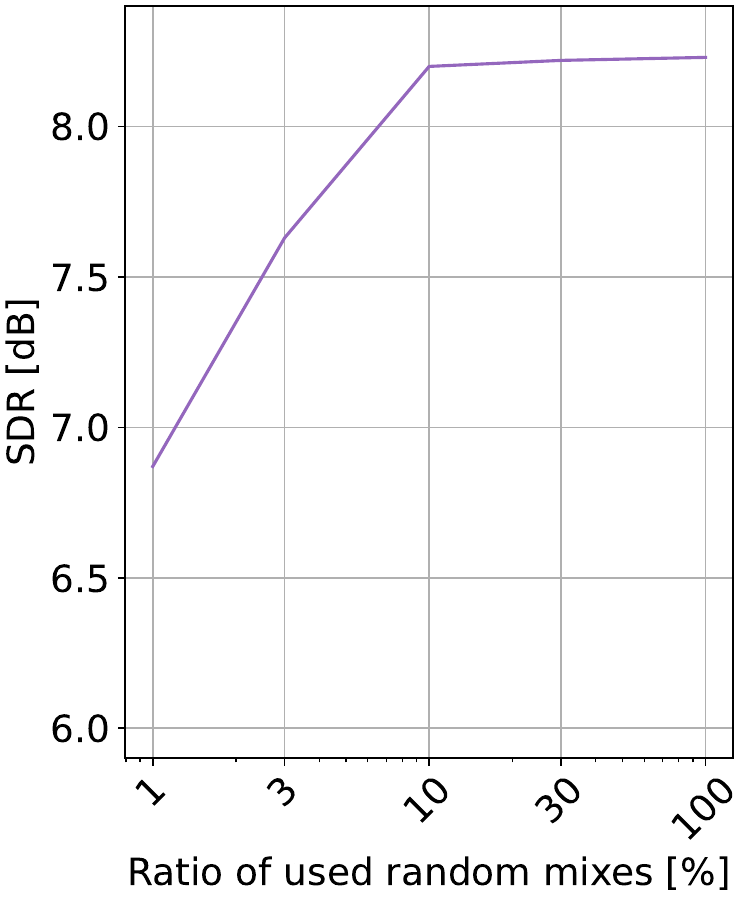}}
  \centerline{(a) With full data
  }\medskip
\end{minipage}
\hfill
\begin{minipage}[b]{0.48\linewidth}
  \centering
  \centerline{\includegraphics[width=4.0cm]{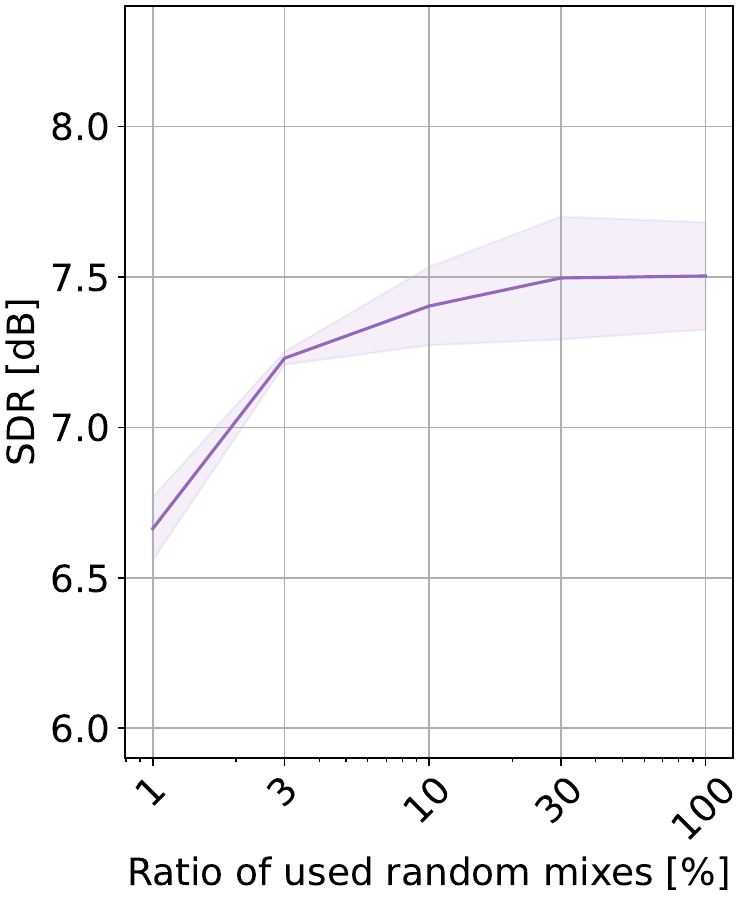}}
  \centerline{(b) With reduced data
  }\medskip
\end{minipage}
\vspace{-0.5cm}
\caption{Test SDR when varying the amount of random mixes created from (a) the full training data and (b) a reduced set of songs. Models are trained by repeatedly sampling batches from this fixed set of random mixes for $I=\text{470k}$ iterations, with the size of this set indicated on the x-axis as a ratio to the total number of training examples.}\label{fig:effective_number}
\vspace{-.35cm}
\end{figure}

In this section, we investigate the amount of random mixes needed for effective training. We first define a  number $R$ of unique random mixes as a fraction of the total number $I\times B$ of training examples (which as discussed previously is \num{1880000}). We then train models by sampling training examples only from those fixed $R$ random mixes, repeating them throughout training
 until we reach $I\times B$ samples,
without using any original mixes.
In Fig.~\ref{fig:effective_number}(a), we see that even when the model is trained with only 10\% of the maximum number of random mixes, there is almost no decrease in performance. This result somewhat explains the performance gain of the $p\!=\!0.9$ model in Fig.~\ref{fig:randmix_ratio}, where \num{188000} random mixes used in conjunction with original mixes were actually enough to train a model with competitive performance. On the other hand, it is also noticeable that the final performance of the $p\!=\!0.9$ model from Fig.~\ref{fig:randmix_ratio} (\SI{7.2}{\dB} SDR) is far below that of the model trained solely by repeating the 10\% of fixed random mixes from Fig.~\ref{fig:effective_number}(a) (\SI{8.2}{\dB}). This likely indicates that the model trained  with $p\!=\!0.9$ original mixes overfits to the original mixes. It is also possible that original mixes are less informative for model training, which we will investigate further in Section \ref{sec:consistency}.

In Fig.~\ref{fig:effective_number}(b), we create the same fixed number of random mixes, but sample training stems from a reduced dataset obtained by randomly selecting about half of the songs from the 86 training songs in MUSDB. We repeat the song selection and random mix creation with 3 different random seeds, where the reduced dataset's length was 2.3, 2.6, and 3.0 hours. 
The shaded regions in the figure show the standard deviation across seeds. Similar to Fig.~\ref{fig:effective_number}(a),
Fig.~\ref{fig:effective_number}(b) also confirms that training with 10\% of fixed random mixes is comparable to using a unique random mix for each training sample (100\%). Furthermore, it is noteworthy that just repeatedly using \num{90} hours of random mixes obtained from 86 songs (the 3\% setting in Fig.~\ref{fig:effective_number}(a)) performs on par or better than using about \num{3000} hours of never-repeating random mixes from the reduced song data (100\% setting in Fig.~\ref{fig:effective_number}(b)). 
It thus appears that a small amount of random mixes composed of stems from a larger set of songs performs better than a large amount of random mixes composed of stems from a smaller set of songs. Thus, the diversity or novelty of random mixes may be more important than their amount; data augmentation alone is not enough, fresh data is necessary for future improvements.


\vspace{-.03cm}
\section{On Consistent Beat and Tonality} \label{sec:consistency}
\begin{figure}[t]

\begin{minipage}[b]{.48\linewidth}
  \centering
  \centerline{\includegraphics[width=4.0cm]{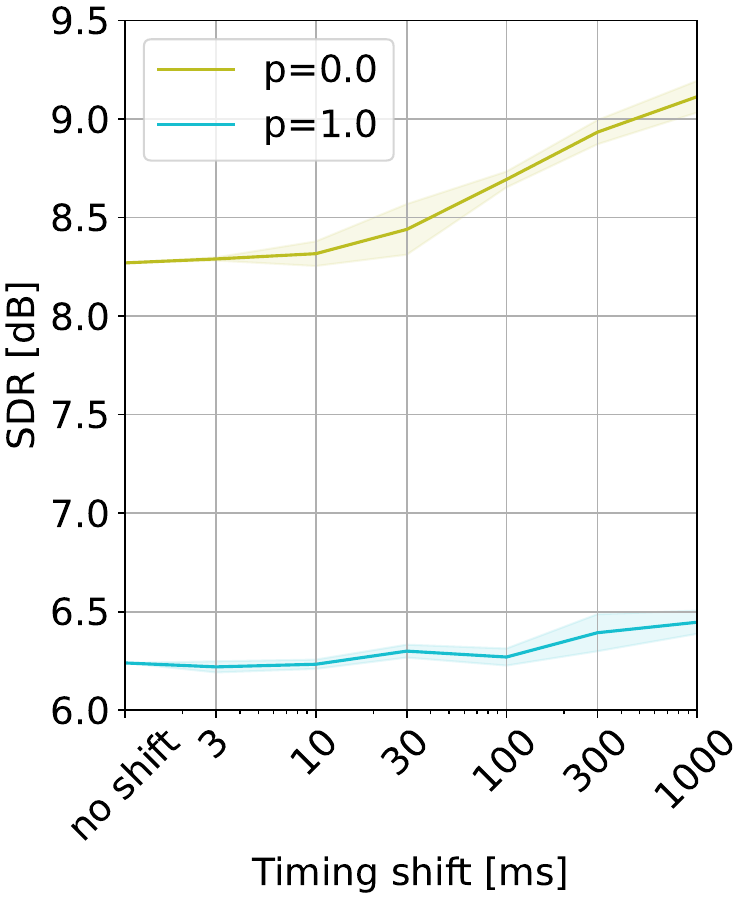}}
  \centerline{(a) Inconsistent beat}\medskip
\end{minipage}
\hfill
\begin{minipage}[b]{0.48\linewidth}
  \centering
  \centerline{\includegraphics[width=4.0cm]{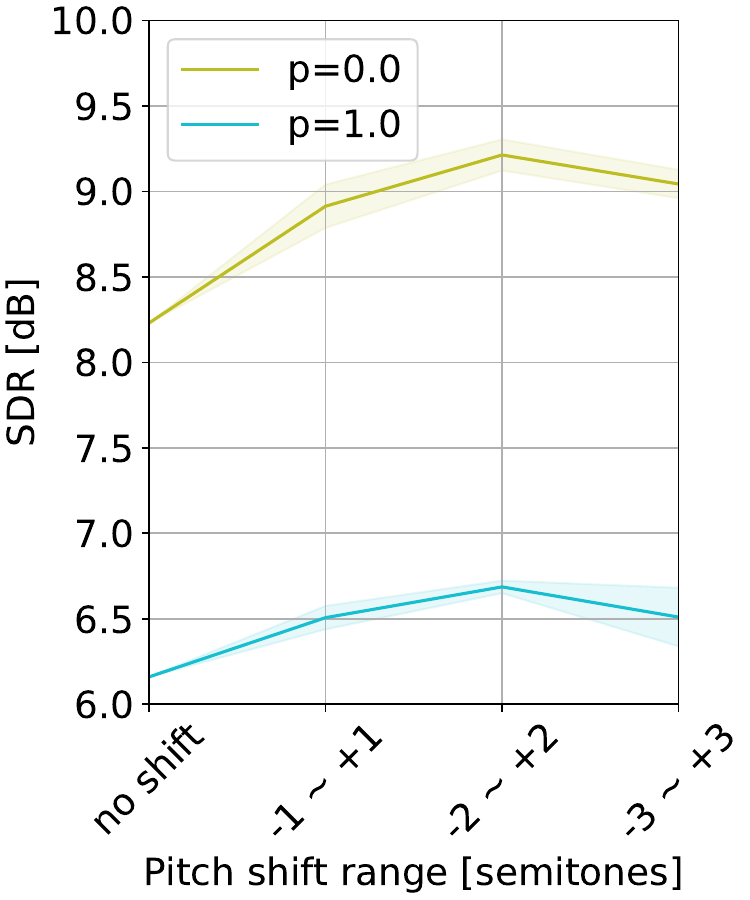}}
  \centerline{(b) Inconsistent tonality}\medskip
\end{minipage}
\vspace{-0.6cm}
\caption{Test SDR for various (a) timing and (b) pitch modified test sets using the models trained with only original ($p\!=\!1.0$) or only random mixes ($p\!=\!0.0$) from Fig.~\ref{fig:randmix_ratio}. Solid lines show the average across test sets generated from 3 different random seeds, and the shaded regions show the standard deviation.}
\label{fig:modified_testset}
\vspace{-.6cm}
\end{figure}

\vspace{-.02cm}
In this section, we analyze how consistent beat and tonality across different stems of mixtures affect performance in music source separation. In Fig.~\ref{fig:modified_testset}, on the test set of MUSDB-HQ, we apply (a) timing shift and (b) pitch shift to each stem to form mixtures that have (a) inconsistent beat and (b) inconsistent tonality, respectively, even as all stems are from the original song. Specifically, timing shift parameters for each stem are sampled from uniform distribution $\mathcal{U}(-t_s,t_s)$, where $t_s$ is the range of timing shift (x-axis in Fig.~\ref{fig:modified_testset}(a)).
Pitch shift values in semitones are discretely sampled up to a maximum shift of 3 semitones up or down, as shown in the x-axis in Fig.~\ref{fig:modified_testset}(b). For each parameter, 3 different random seeds are used. Then, we separate the mixtures using the $p\!=\!0.0$ and $p\!=\!1.0$ models from Fig.~\ref{fig:randmix_ratio}, i.e., models trained with only random and only original mixes, respectively.

In Fig.~\ref{fig:modified_testset}(a), we observe that test sets with inconsistent beat are easier to separate (i.e., higher SDR). Even when timing shift parameters are smaller than STFT parameters (e.g., a shift of $t_s=\SI{30}{\milli\second}$ is smaller than both the FFT size (\SI{186}{\milli\second}) and hop size (\SI{46}{\milli\second}) of the model), performance starts to increase. When $t_s=\SI{1}{\second}$, the performance gain is almost \SI{0.9}{\dB}. We also confirm that inconsistent tonality from pitch shifting makes source separation much easier in Fig.~\ref{fig:modified_testset}(b), as we observe improvements up to \SI{1.0}{\dB}.

\begin{figure}[t]

\begin{minipage}[b]{.48\linewidth}
  \centering
  \centerline{\includegraphics[width=4.0cm]{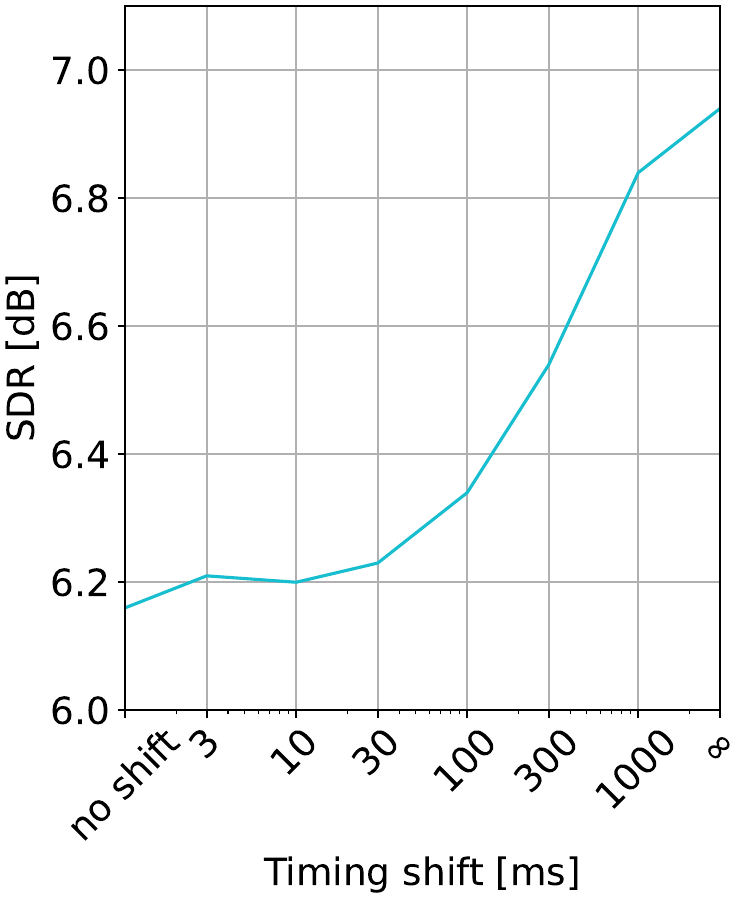}}
  \centerline{(a) Inconsistent beat}\medskip
\end{minipage}
\hfill
\begin{minipage}[b]{0.48\linewidth}
  \centering
  \centerline{\includegraphics[width=4.0cm]{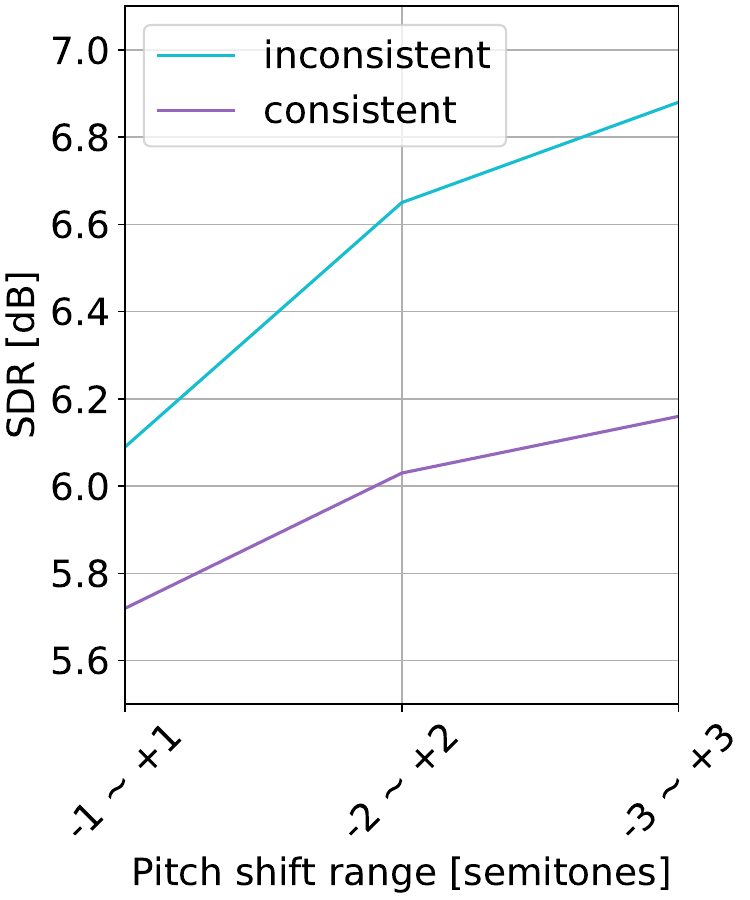}}
  \centerline{(b) Inconsistent tonality}\medskip
\end{minipage}
\vspace{-0.6cm}
\caption{Test SDR of models trained on original mixes with (a) slightly shifted timing or (b) inconsistent tonality across stems.}
\label{fig:shift_training}
\vspace{-.5cm}
\end{figure}

Furthermore, we investigate the effects of timing and pitch shift during training, but evaluated on the unmodified MUSDB-HQ test set. In Fig.~\ref{fig:shift_training}(a), we train models without random mixing but with slight time shifts applied to each stem during training, and add the $t_s\!=\!\infty$ case, where 6-second training samples can come from anywhere in the same song, i.e., within-song random mixing. In Fig.~\ref{fig:shift_training}(a), we observe that for shifts greater than \SI{100}{\milli\second}, which is longer than the STFT hop size, the scores of the models start to increase, reaching \SI{6.8}{\dB} SDR at $t_s\!=\!\SI{1}{\second}$ shift, which is only \SI{0.1}{\dB} lower than the $t_s\!=\!\infty$ shift. 

Fig.~\ref{fig:shift_training}(b) displays results after training with only original mixes, but applying pitch shifts to each stem in both a `consistent' (same pitch shift applied to each stem, the augmentation used in other experiments) and `inconsistent'  (different pitch shift applied to each stem) manner. %
Scores increase with larger amounts of pitch shift for both cases, with inconsistent pitch shift providing a notable improvement.
The best consistent pitch shifting (\SI{6.2}{\dB}), which corresponds to the $p\!=\!1.0$ model from Fig.~\ref{fig:randmix_ratio}, is much worse than the best inconsistent one, suggesting that inconsistent tonality between stems during training is an important component of random mixing.
Still, we note that even the best-scoring configuration of within-song random mixing in Fig.~\ref{fig:shift_training} scores clearly worse than full random mixing. %

In Table \ref{tab:randomly_made}, to further investigate how domain mismatch affects the performance of MSS networks, we randomly choose stems from different songs in the train set of MUSDB-HQ and mix them to make a dataset of 86 beat- and tonality-inconsistent songs from which training chunks will be sampled (until now, random mixing was at the 6-second chunk level, resulting in many more possibilities). 
For each song, the four chosen stems are trimmed to match the shortest one.
This results in new datasets across 3 random seeds that each have length of \num{2.7} hours. Then, we train models on the random songs without random mixing (similarly to the $p\!=\!1.0$ model in Fig.~\ref{fig:randmix_ratio}) and compare them to models trained on
the original songs. 
For fair comparison, we used 3 random \num{2.7} hour subsets obtained by randomly trimming each original song. %

Surprisingly, %
data with inconsistent beat and tonality (\textit{Random}) leads to significant performance degradation compared to consistent beat and tonality (\textit{Original}), dropping from \SI{6.1}{\dB} to \SI{0.7}{\dB} SDR. 
In contrast, a model trained using \num{2.7} hours of random mixes obtained at the 6-second chunk level ($R=0.09\%$ in Section~\ref{sec:effective}), where the training examples are likely much more diverse, obtained \SI{5.0}{\dB} SDR.
This implies that the domain consistency matters in music source separation, all the more so as data diversity becomes limited. 
This could explain why, in the task of vocal harmony separation where consistency is arguably even stronger, random mixing was found to be detrimental on a small dataset (104 minutes) in \cite{sarkar2021vocal}, but significantly beneficial on a much larger one (400 hours) in \cite{jeon2023medleyvox}.

\begin{table}[t]
\sisetup{
detect-weight, %
mode=text, %
tight-spacing=true,
round-mode=places,
round-precision=1,
table-format=3.1,
table-number-alignment=center
}
\caption{Test SDR (mean $\pm$ standard deviation in [dB]) for models trained with original (consistent beat and tonality) and randomly made (inconsistent beat and tonality) data at the song level (not for every chunk). 
}
\label{tab:randomly_made}
\vspace{.05cm}
\setlength{\tabcolsep}{2pt}
\resizebox{\columnwidth}{!}{
\begin{tabular}{l*{5}{S@{\,\( \pm \)\,}S[table-format=1.1]}}
\toprule
Training data & \multicolumn{2}{c}{vocals}                  & \multicolumn{2}{c}{bass}                  & \multicolumn{2}{c}{drums}      & \multicolumn{2}{c}{other}      & \multicolumn{2}{c}{avg}      \\ \midrule
\textit{Original}      &  7.42 & 0.07   & 5.43 & 0.14 & 6.64 & 0.07 & 4.72 & 0.06 & 6.05 & 0.06 \\
\textit{Random}        & -0.86 & 0.89             & -1.44 & 1.02           & 2.20 & 2.17 & 2.96 & 0.61 & 0.72 & 1.09  \\ \bottomrule
\end{tabular}
}
\vspace{-.6cm}
\end{table}

\vspace{-.3cm}
\section{Conclusions and Future Work}
\vspace{-.3cm}

In this paper, we took steps towards explaining the effectiveness of random mixing, an obviously non-musical data augmentation technique that has become indispensable in music source separation research.
Our results illustrated that both data diversity and domain consistency in terms of beat and tonality are important for music source separation, 
but the data diversity provided by random mixing seems more beneficial unless the amount of random mixes is strictly restricted.
We also empirically demonstrated that actual performance gain of random mixing comes from the diversity of songs, not solely from the infinitely many combinations of random mixes.
Furthermore, it was confirmed that the inconsistent beat and tonality of random mixes make separation easier, which perhaps leads to a stronger learning signal during network training. 
In the future, we plan to validate our results on larger recently released datasets~\cite{pereira2023moisesdb} and explore learning curricula based on random and original mixes.

\balance
\bibliographystyle{IEEEtran}
\bibliography{strings,refs}

\end{document}